# Anisotropy of Interfacial Energy in Five Dimensions

**Authors:**

Vasily V. Bulatov

Lawrence Livermore National Laboratory, Livermore, California 94550, USA

Bryan W. Reed

Lawrence Livermore National Laboratory, Livermore, California 94550, USA

Mukul Kumar

Lawrence Livermore National Laboratory, Livermore, California 94550, USA

**Abstract:**

Anisotropy of interfacial energy is the principal driving force for material microstructure evolution yet its origins remain uncertain and a quantitative description lacking. We present and justify a concise hypothesis on the topography and topology of the functional space of grain boundary energies and, based on this hypothesis, construct a closed-form function that quantitatively describes energy variations in the entire 5-space of macroscopic parameters defining grain boundary geometry. The new function is found to be universal for the crystallography class of face-centered cubic metals.

**Authors' contact information:**

| | |
|---|---|
| Vasily V. Bulatov | The corresponding author |
| E-mail: | bulatov1@llnl.gov |
| Postal address: | LLNL, 7000 East Avenue, Livermore, CA 94550, USA |
| Phone: | 925-423-0124 |
| Fax: | 925-423-0785 |
| Bryan W. Reed | |
| E-mail: | reed12@llnl.gov |
| Postal address: | LLNL, 7000 East Avenue, Livermore, CA 94550, USA |
| Phone: | 925-423-3617 |
| Fax: | 925-423-7040 |
| Mukul Kumar | |
| E-mail: | kumar3@llnl.gov |
| Postal address: | LLNL, 7000 East Avenue, Livermore, CA 94550, USA |
| Phone: | 925-422-0600 |
| Fax: | 925-424-4737 |

Grain boundaries (GB) affect a great many physical, chemical and mechanical properties of crystalline solids including electrical and thermal conductivity, thermal coarsening, corrosion resistance, impurity segregation, hydrogen embrittlement, stress corrosion cracking, and mechanical strength and ductility [1]. The specific contribution of each given boundary to the material properties is largely defined by its excess energy per unit area that depends primarily on boundary geometry. The geometric character of a boundary is defined by five macroscopic degrees of freedom (DOFs) that allow several alternative representations the most common of which is in terms of grain misorientation (three DOFs) plus boundary plane inclination (two more DOFs) [2,3,4]. The GB energy is known to be anisotropic and can vary significantly with both the misorientation and the inclination, especially in face-centered cubic (FCC) metals [5,6,7].

Despite its recognized importance and an extensive body of literature on the subject, the GB energy anisotropy has not been systematically quantified. Particularly lacking is a sense of the GB energy dependence on plane inclination, which is difficult to quantify experimentally. Reflecting this poor understanding and with very few exceptions current models of GB network evolution disregard GB energy anisotropy altogether or, at best, account only for the misorientation dependence of the GB energy. Yet, depending on the GB plane orientation, boundaries of the same grain misorientation can have vastly different energies [6,7]. Furthermore, except for nano crystals in which the grains can rotate, it is precisely the variations of the GB energy with plane inclination – and not misorientation – that define the capillary force driving boundary motion.

Previous attempts to relate the GB energy anisotropy to lattice geometry have been unsuccessful raising legitimate doubts about the very existence of a concise relationship [8]. The primary difficulty stems from the high dimensionality and intricate topology of the "misorientation + inclination" 5-space making it difficult to decouple the inclination-dependent variations from the anisotropy in the misorientation subspace. At the same time, experimental observations and atomistic calculations reveal certain common and consistent trends in GB energy variations among materials of a given crystallography class (FCC, BCC, etc.) suggesting that lattice geometry does play a role in defining the GB energy anisotropy. Particularly noteworthy are observed close correlations between the GB energy and the spacing between lattice planes parallel to the boundary and the magnitude of the planar unit cell area [9,10,11,12]. Yet the strongest impetus to continue to seek a purely geometric description of the GB energy anisotropy was a recent observation of a close scaling among the energies of 388 geometrically distinct boundaries computed for four FCC metals Ni, Al, Au and Cu [6,7]. The observed scaling suggests that a functional relationship between the GB energy and five geometric DOFs does exist but we do not know what it is.

Existing geometric models relate the GB energy anisotropy to a measure of lattice coincidence [13], plane inter-locking [14], hard-sphere packing [15], excess free volume [9], boundary proximity to symmetric cusp orientations [10,16] and others. While these models capture interfacial energy variations within limited sub-sets of the 5-space, all our attempts so far to extend the same models to the entire sample of 388 boundary geometries have failed prompting us to consider an entirely different approach. Following [6,7] we assume that the GB energy is a continuous function of five macroscopic DOFs and, thus, the energy of a given boundary can be approximated by interpolation between nearby boundaries with known energies. By itself this proposition is hardly constructive given that many more than 388 samples are needed in order to enable accurate energy interpolation in the entire 5-space ($3^5 \approx 388$). To make interpolation practical it is useful to first grasp the global topography of the energy function in the 5-space. Here we develop an accurate closed-form expression describing variations of interfacial energy in face-centered cubic (FCC) metals as a function of GB geometry. Our approach relates interfacial energy anisotropy to the global topography and connectivity (topology) of the 5-space dominated by special low-dimensional subsets termed *grofs*. The new interfacial energy function uses grofs as scaffolding for hierarchical interpolation providing an accurate description of GB energy anisotropy in four FCC metals in the entire 5-space. Among the numerical parameters defining the GB energy function, only two are found to be metal-specific with the rest held universal for the FCC crystallography class.

As a motivation, consider a much simpler yet pertinent example of energy anisotropy among crystal interfaces, namely the variations of surface energy as a function of surface orientation. A fairly accurate model of surface energy anisotropy is obtained by counting the density of first nearest-neighbor (1NN) bonds cut by the surface plane [17,18]. A polar plot of the broken 1NN bond density is given in figure 1. Defining the topography of the surface energy variations are six grooves crisscrossing the otherwise smooth energy function. Each groove corresponds to a 1d-set of planes that do not cut one of the six 1NN bonds in the FCC crystal. Orientations for which two or more grooves intersect correspond to cusps: that all cusps seen on the figure lie at the intersection of exactly two or three grooves, in $\langle 100 \rangle$ and $\langle 111 \rangle$ directions, respectively, is a consequence of the symmetry of the FCC lattice.

Our hypothesis is that, just like the surface energy in figure 1, the topography of GB energy function is defined by grooves, i.e. special sub-sets of the 5-space where the energy is locally minimal with respect to variations locally orthogonal to the set. To avoid confusion with the real physical grooves observed at the intersections of GB with crystal surfaces, we refer to such locally minimal subsets of 5-orientations as *grofs* (from Old Norse gróf for 'pit'). A grof of order k or, simply, k-grof of a function in the Nd-space is defined here as a contiguous (N-k)-dimensional subset of the N-space such that most points in the subset possess the following *grof property*: the function is smooth in the (N-k)-subset itself but is a cusp in the complementary k-subspace.

In the case of surface energy shown in figure 1, there are six distinct 1-grofs and 14 distinct 2-grofs on the 1NN function. In this 2d case, each 2-grof is a true cusp since its dimensionality is zero. Similarly, a 5-grof in the 5-space of GB orientations is a true cusp. In defining the grofs we stated that most but not all points in a grof set must possess the grof property. This leaves room for possible intersections of two or more grofs. In the 5-space, grof intersections can be grofs of higher order k but not necessarily 5-grofs (true cusps). The grofs are important since, as low-dimensional subsets of orientations with (locally) minimal energies, they define the topography of the energy function in the 5-space. Topology (connectivity) of grofs is defined by crystal symmetries.

If real, the proposed grof topography should greatly simplify interpolation in the 5-space by allowing, at least in principle, to approximate the energies of various boundaries by their proximity to grofs. As a first step, we used the datasets of 388 computed GB energies to identify important grofs of low energy and/or high symmetry. Precise conditions defining energy grofs in the 5-space remain uncertain. On one hand, the excess energy must account for bond distortion related to the degree of lattice coincidence between the two grains (lattice coincidence is commonly quantified by the inverse Σ–number, or $Σ^{-1}$[1,13]). On the other hand, the energy must also depend on the area density of bonds cut by the interface and, thus, on the boundary plane orientation.

Based on the GB energy data presently available, we hypothesize that many rational grain misorientations of high lattice coincidence are 3-grofs (cuspy in the 3-subspace of grain rotations but smooth in the 2-subspace of plane inclinations). Of these, Σ1, Σ3, Σ5, Σ11, and Σ17 grofs are partially accounted for in the resulting 5d function, whereas all other 3-grofs of this type are deemed too shallow and ignored. Some 1d subsets of *twist boundaries* (for which the rotation axis is perpendicular to the boundary plane) appear to be 4-grofs. In particular, excess energies of all ⟨111⟩-twist boundaries are low because a twist rotation around any of the four ⟨111⟩ axes leaves three 1NN ⟨110⟩ bonds intact (in the boundary plane). Likewise, twist rotations around ⟨110⟩ and ⟨100⟩ axes leave intact one or two 1NN bonds. Tellingly, the mentioned subsets of symmetric twist boundaries correspond to the three most widely spaced {111}, {100}, and {110} boundary planes with well-documented low energies [10,11,12]. These three high-symmetry 4-grofs are also accounted for in the 5d function. Some of the grofs intersect. For example, a coherent twin boundary is simultaneously a Σ3 boundary and a ⟨111⟩-twist boundary: this boundary is a true cusp (5-grof).

However impressive, the set of 388 boundaries examined in [6,7] was selected based on criteria favoring high symmetry boundaries of high lattice coincidence (low Σ) and/or with low index boundary planes. Such selection rules generally disfavor GBs vicinal (adjacent) to the high-symmetry subsets and other low-symmetry boundaries, leaving attribution of boundary subsets as grofs somewhat uncertain. At the same time, many of the high-symmetry low-dimensional subsets in the 5-space, including the above mentioned 3- and 4-grofs, are well sampled. Leaving precise attribution of grofs for future work, our present interpolation approach is: (a) quantify and fit the GB energy variations within several well-sampled low-dimensional subsets (preferably grofs) of the 5-space and (b) use the so-fitted subsets as scaffolding onto which the global 5-

dimensional interpolation function is stretched. For constructing our interpolation function we opted to use three rotational subsets of highest symmetry: the ⟨100⟩, the ⟨111⟩ and the ⟨110⟩ rotation sets. Here, by a rotation set we mean all possible boundaries that can be produced by a rotation around the set axis. Even though it remains uncertain whether the three selected scaffolding 3d-sets are actually grofs, some of their low-dimensional subsets are nearly certainly grofs. Whatever the case, our particular choice is justified, postfactum, by observing that variations of GB energy within each of the three selected scaffolding sets are simple.

A complete definition of the 5d function is presented online in the supplementary data that also includes a MATLAB® code for the function itself. Here we only briefly describe how it is constructed. First we accurately parameterize variations of the GB energy within the three rotational 3d-sets. The approximating 3d-sets are parameterized hierarchically: first in their special 1d sub-sets, then in their 2d sub-sets and finally the entire 3d-sets. As illustrated in figure 2, the lower dimensional sub-sets serve as scaffolding for the higher dimensional sub-sets. Once fully parameterized, three rotational sets provide a 3d scaffolding for energy interpolation in the entire 5-space: the energy of a boundary that is not in any of the three scaffolding sets is approximated by the boundary's proximity to the sets. Interpolation can be accurate provided no boundary in the 5-space is too far away from all three scaffolding sets that criss-cross the 5-space. Fortunately, the high symmetry of the cubic lattice enhances the coverage of the 5-space by the three 3d sets. Interpolation between the scaffolding sets requires a measure of distance between the boundaries - a metric in the 5-space. As detailed in the SM, we circumvent the notorious uncertainty in defining such a metric [19,20,21] by using two sub-metrics to evaluate the proximity: one sub-metric in misorientation 3-space and another sub-metric in the 2-space of boundary plane inclinations.

Initially we fitted the 43 numerical parameters defining our interpolation function to the energies of 388 boundaries for each of the four metals separately. We observed that: (a) each of the resulting four interpolation functions reproduces its corresponding energy dataset to within RMS error of 5%; (b) the overall scale of energy variations is not correlated with the other fitting parameters that define the shape of the GB energy function; (c) within errors derived from the $\chi^2$ analysis, 27 of 42 fitting parameters are the same for all four elements; (d) statistically significant inter-element variations in the remaining 15 parameters are linearly dependent and can be accurately described with just one degree of freedom. These insights prompted us to re-define our interpolation function as universal, with just two material-specific parameters, and re-fit it to all 388x4=1552 available boundary energies. The resulting fit is shown in figure 3. That it is possible to capture the global 5d topography of the GB energy using a universal function with just two metal-specific parameters is a remarkable reduction in complexity. To the extent that the proposed energy function is indeed universal, one needs to measure or compute just two GB energies, e.g. that of the $\Sigma 3$ coherent twin boundary and one of the high energy random boundaries, to fully define the GB energy anisotropy for a given FCC material in the entire 5-space.

In summary, considering that most of the numerical parameters in our GB energy function are the same for all four FCC metals and only two parameters are material-specific, our answer to the question "what matters more - lattice geometry or details of atomic structure and bonding across the interface?" is - both matter, but not in equal measures.

This work was performed under the auspices of the US Department of Energy by Lawrence Livermore National Laboratory under Contract W-7405-Eng-48. The authors were supported by the US DOE Office of Basic Energy Sciences, Division of Materials Sciences and Engineering. We are indebted to D. Olmsted for many useful discussions and to L. Zepeda-Ruiz, T. Oppelstrup, and J. Mason for their advice on various technical aspects of this work.

Supplementary data associated with this article can be found, in the online version, at http://xx.xxx.xxx.

**Figures captions:**

**Figure 1**: Density of broken 1NN bonds as a function of plane orientation.

**Figure 2:** GB energy variations in low-dimensional subsets within the ⟨111⟩ rotational set. The symbols are energies taken from the datasets of 388 boundaries with the assumed error bars of 5% while the lines and the surface show the fit function itself. (**A**) Energies of twist boundaries as a function of rotation angle. (**B**) Energies of symmetric tilt boundaries as a function of rotation angle. (**C**) Energies of all tilt (symmetric + asymmetric) boundaries as a function of rotation angle $\xi$ and asymmetry angle $\eta$ (defined in SM) in Ni. Boundary character – twist, tilt or mixed - is defined by the orientation of the boundary plane relative to the rotation axis: twist/tilt boundaries have their planes perpendicular/parallel to the rotation axis and all other boundaries are considered mixed. Tilt boundaries are further classified as symmetric if the lattice planes parallel to the boundary are the same in both crystal grains and asymmetric otherwise. The lines at two borders of the surface at $\eta=0º$ and $\eta=120º$ in (C) are the same as the blue line in (B).

**Figure 3:** The atomistic energies of all 388 boundaries in all four elements reported in (6,7) plotted against the energies of the same boundaries computed from the interpolation function. The solid line corresponds to a perfect fit, while the dashed lines indicate the assumed 5% error bounds. For each metal the energies are scaled by the maximum energy computed over all 388 boundaries for the same metal.

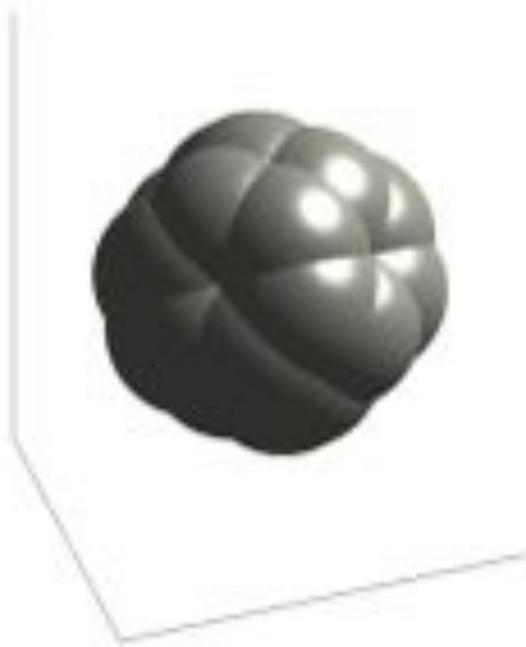

**Figure 1**: Density of broken 1NN bonds as a function of plane orientation.

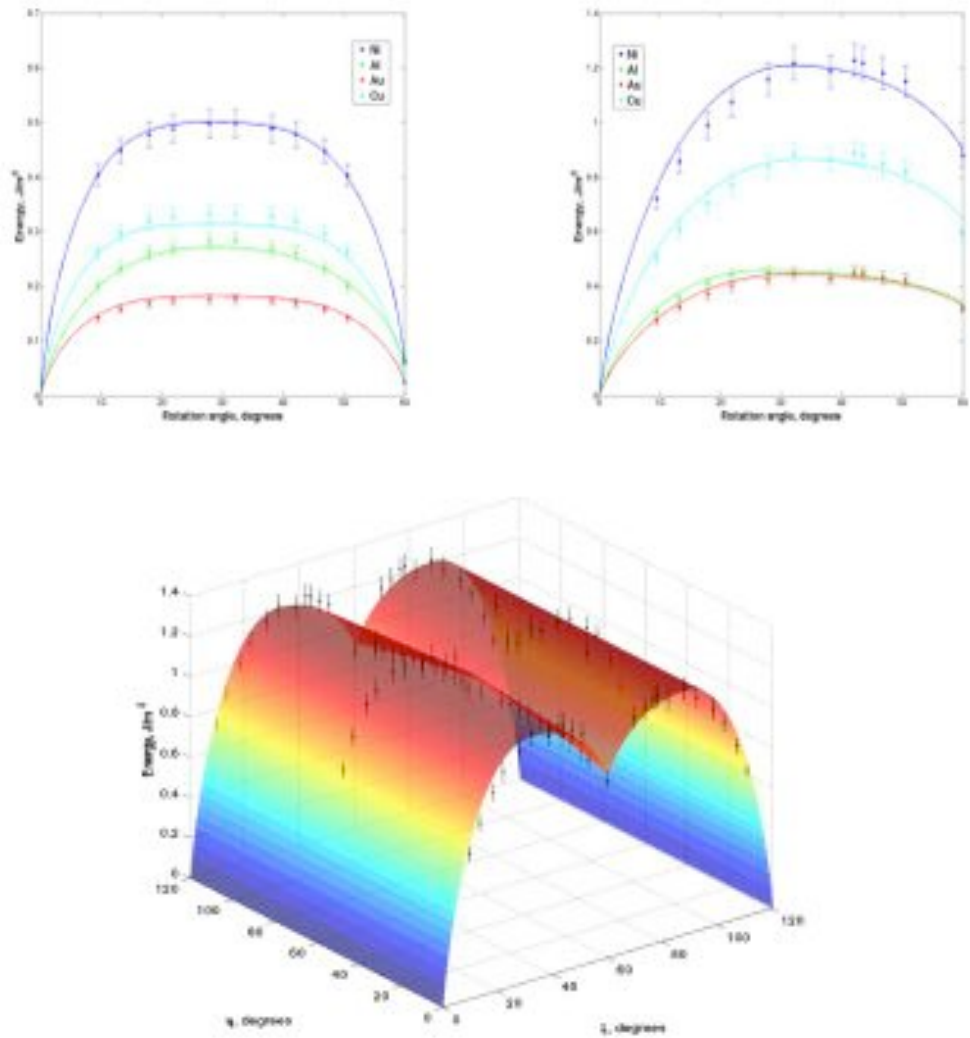

**Figure 2:** GB energy variations in low-dimensional subsets within the ⟨111⟩ rotational set. The symbols are energies taken from the datasets of 388 boundaries with the assumed error bars of 5% while the lines and the surface show the fit function itself. (**A**) Energies of twist boundaries as a function of rotation angle. (**B**) Energies of symmetric tilt boundaries as a function of rotation angle. (**C**) Energies of all tilt (symmetric + asymmetric) boundaries as a function of rotation angle ξ and asymmetry angle η (defined in SM) in Ni. Boundary character – twist, tilt or mixed - is defined by the orientation of the boundary plane relative to the rotation axis: twist/tilt boundaries have their planes perpendicular/parallel to the rotation axis and all other boundaries are considered mixed. Tilt boundaries are further classified as symmetric if the lattice planes parallel to the boundary are the same in both crystal grains and asymmetric otherwise. The lines at two borders of the surface at η=0° and η=120° in (C) are the same as the blue line in (B).

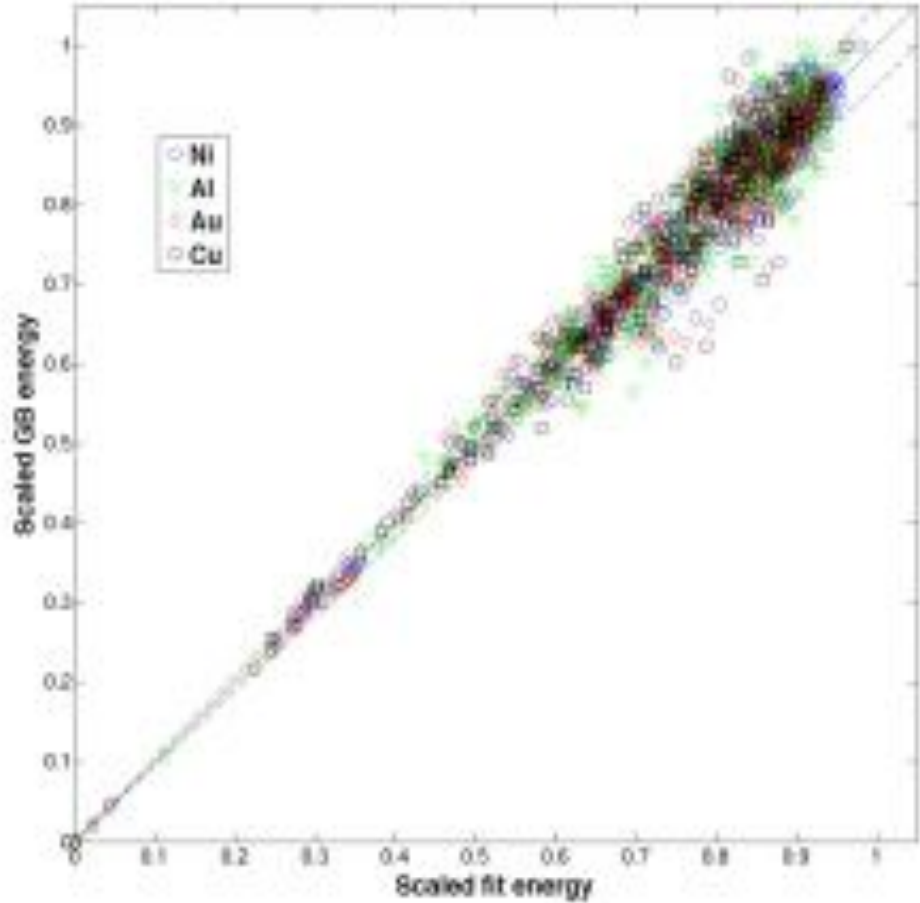

**Figure 3:** The atomistic energies of all 388 boundaries in all four elements reported in (6,7) plotted against the energies of the same boundaries computed from the interpolation function. The solid line corresponds to a perfect fit, while the dashed lines indicate the assumed 5% error bounds. For each metal the energies are scaled by the maximum energy computed over all 388 boundaries for the same metal.

Supplementary data for

# Anisotropy of Interfacial Energy in Five Dimensions

Vasily V. Bulatov*, Bryan W. Reed and Mukul Kumar

Lawrence Livermore National Laboratory, Livermore, California 94550, USA

All correspondence should be addressed to:  bulatov1@llnl.gov

# 1. Motivation

Figure S1 presents a comparison of the grain boundary energies for four FCC metals calculated in previous work [s1,s2] using appropriate atomistic models. Based on data presented in this plot and in [s2] we make several preliminary observations mirroring those in [s1,s2] and relevant for the construction of our energy function. First, the scaled energies in Ni and Au are practically indistinguishable to within random scatter with a root-mean-square (RMS) value of order 5%. Thus, to this precision, most of the statistically reliable information distinguishing these two elements is accounted for by an overall energy scale factor (henceforth called $\epsilon_{RGB}$, the energy of a hypothetical random grain boundary). Second, the other two elements (Al and Cu) show very similar scaled energies, again allowing 5% random scatter mostly for high-energy boundaries. Third, most systematic differences among the scaled energies concern a relatively small number of boundaries with low scaled energies, for which Cu exhibits somewhat lower energies and Al exhibits considerable higher energies than Ni. The most obvious outliers are the $\Sigma 3$ boundaries in Al. Taken together, these observations suggest that it may be possible to capture much of the differences in scaled energies with a small number of element-specific parameters. This is borne out in the present work where we show that, in addition to $\epsilon_{RGB}$, a single parameter $\epsilon_{twin}$ representing the scaled energy of a $\Sigma 3$ twin boundary suffices to capture the inter-element variation to an RMS scatter of 5%.

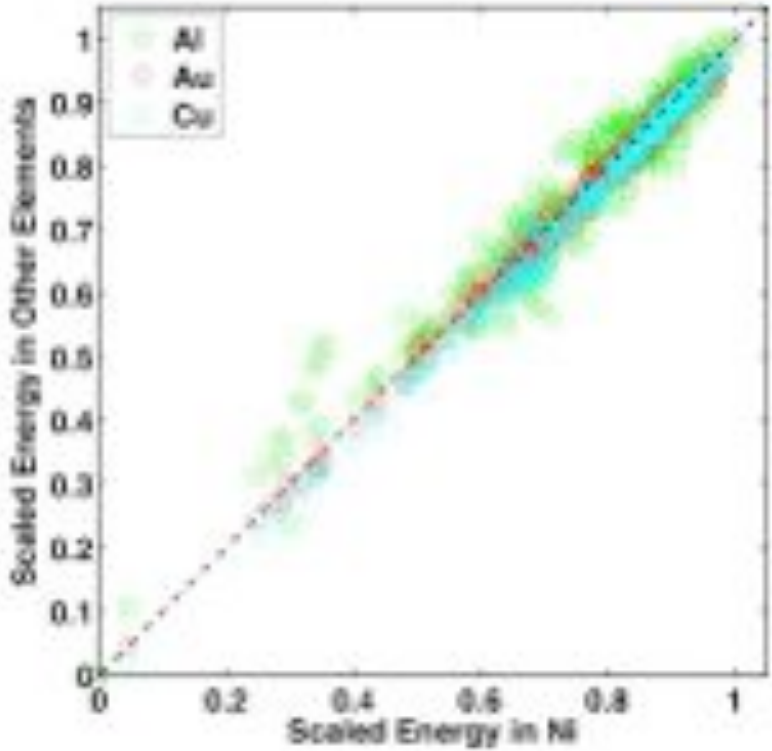

**Figure S1:** Grain boundary energies computed atomistically in [s1,s2] for four FCC metals. The set of 388 energies for each element was scaled as a fraction of the maximum energy for that element.

## 2. Detailed specification of the energy function

### 2.1 General concept of hierarchical interpolation

The energy function is based on interpolation between three special subsets of the five-dimensional space (or simply 5-space) that include all boundaries obtainable by rotations about <100>, <110>, and <111> axes (i.e. the <100>, <110>, and <111> "sets"). The interpolation is hierarchical, building up the full function of five variables from one-, two-, and three-dimensional subsets. Each of the three selected scaffolding "sets" is a three-dimensional subset of the full 5-space: any boundary in a given 3-set is fully defined by the angle $\xi$ of rotation around the set axis (say [100]) and two additional angles $\eta$ and $\phi$ (defined below) that fix the boundary plane inclination. Each 3-set contains sub-sets of still lower dimensions, such as the subset of pure-twist boundaries for which the misorientation axis is normal to the boundary plane. This latter subset is one-dimensional, fully defined by the rotation angle $\xi$. Tilt boundaries are defined as a subset of boundaries for which the rotation axis lies in the plane of the boundary: in this case two degrees of freedom $\xi$ and $\eta$ are needed and used to specify a tilt boundary within a given 3-set. The two-dimensional subset of tilt boundaries is further subdivided into symmetric tilt grain boundaries (STGB), with $\eta = 0$ (for such boundary the boundary plane is a plane of mirror symmetry), and asymmetric tilt grain boundaries (ATGB), with $\eta \neq 0$. The energies of boundaries within the two one-dimensional subsets (twist boundaries and STGBs) are explicitly curve-fit for each 3-set. Then the energies of boundaries in the two-dimensional subsets of ATGBs are interpolated from their corresponding one-dimensional subsets of STGBs. Finally, boundaries of mixed characters (neither twist nor tilt) in each set are further interpolated between their ATGB and pure-twist components. Once the three 3-sets are fully parameterized, the general boundaries with rotation axes other than <100>, <110>, and <111>, are approximated as weighted averages of the energies of the nearest elements in the three scaffolding 3-sets.

   To enable interpolation, we need to define distance, i.e. a metric, in the 5-space. A variety of equally useable metrics can be defined, separately, for the 3-subspace of grain misorientations ($d_3$) [s3] and for the 2-subspace of plane inclinations ($d_2$). For $d_3$ we follow the definition given in [s4] to be re-written below in a different form more convenient for our purposes. Our choice for $d_2$ will be also described in the following. Even though both $d_3$ and $d_2$ are well defined separately, it is far from obvious how to combine any two of such sub-metrics into a measure of distance in the entire 5-space [s4,s5,s6]. To avoid ambiguity in mixing the two sub-metrics, we regard the sub-metric $d_3$ as the primary measure of distance with $d_2$ being secondary. Accordingly, for a given boundary $A$ (with misorientation $R$) and for a given high-symmetry rotation axis [$hkl$] (of type <100>, <110>, or <111>), we first find the rotation $R'$ about [$hkl$] that minimizes the distance $d_3$ between $R'$ and the actual grain misorientation $R$. Then, among all boundaries with misorientation $R'$ we find the boundary $B$ that is nearest to $A$ in the inclination 2-subspace (i.e. that minimizes $d_2$): this specifies a boundary $B$ that is the closest match of $A$ in one of the considered high-symmetry 3-sets <$hkl$>. Thus, we factorize the task of finding boundaries nearest to boundary A: for each scaffolding 3-set we first find which rotation $\xi$ most closely matches A in the misorientation subspace and then find which particular boundary of misorientation $\xi$ is the closest match of A in the inclination sub-space. We use the energies of the so-matched boundaries as approximations (interpolants) to produce an estimate of the energy of boundary A.

The above matching procedure is performed for each of the four <111> axes, three <100> axes, and six <110> axes, and for each of the 24 symmetry-related representations of boundary *A*. The resulting list of 13x24 approximating boundaries *B* is culled to remove redundant symmetry-equivalent variations along with any boundaries *B* with distances $d_3$ to *A* exceeding some pre-defined cutoff distance $d_{hkl}^{max}$. The final boundary energy $\epsilon$ is then taken to be a weighted average of the remaining $\epsilon_{hkl}(B)$ values, with the weighting being a function of the associated $d_3$ values.

## 2.2 Geometrical definitions

With the general interpolation approach established, we now proceed to give precise definitions of the geometrical parameters (Fig. S2(a)). A misorientation is represented in the reference frame of one of the grains as a rotation matrix *R* or, equivalently, as an axis-angle pair (*y*, *a*), which rotates the <100> directions in grain 1 to the <100> directions in grain 2. Because of symmetry, *R* is actually just one representative of an entire coset $R_i = S_i R$, where the $S_i$ are the 24 rotational symmetries of FCC crystals. The boundary is also defined by the boundary plane inclination given by its unit normal vectors $n_1$ and $n_2$ in the reference frames of grains 1 and 2, respectively, related by $n_2 = R_i n_1$.

We now wish to approximate boundary A by a special boundary B obtainable by rotation around rotation axis $A_j$ that can be one of three distinct <100> axes or one of four distinct <111> axes or one of six distinct <110> axes. The approximating boundary B is characterized by its axis-angle pair (*x*, $A_j$) (Fig. S4(b)) or, equivalently, rotation matrix *R'* and the unit normals $m_1$ and $m_2$ defining the boundary plane, with $m_2 = R' m_1$.

To find boundary B that most closely matches (approximates) boundary A we need to define a measure of distance $d_3$ between two rotations *R(y)* and *R'(x)*. A mathematically convenient definition, which also accords with previous usage (21), is $d_3 = 2\sin(d/2)$, where *d* is the angle of the rotation $R^{-1}(x)R(y)$, minimized as a function of *x*. This is schematically shown as a vector sum in Fig. S2(b), though of course this is only a schematic since the product of two rotations is non-commutative and does not follow the rules of vector addition. Straightforward algebra yields

$$d_3 = 2\sin\left(\frac{\psi}{2}\right)\left(1 - (\boldsymbol{a} \cdot \boldsymbol{A_j})^2\right)^{\frac{1}{2}}$$

and

$$\tan\left(\frac{\xi}{2}\right) = (\boldsymbol{a} \cdot \boldsymbol{A_j})\tan\left(\frac{\psi}{2}\right)$$

at the closest point. Having selected a rotation for boundary B best matching A in the misorientation sub-space, we still need to find the best matching inclination. For that we minimize the inclination distance defined as

$$d_2 = 2 - \boldsymbol{m_1} \cdot \boldsymbol{n_1} - \boldsymbol{m_2} \cdot \boldsymbol{n_2}$$

which for small angular deviations is approximately equal to the mean squared angular deviation between the normal vectors $n_i$ of boundary A and the normal vectors $m_i$ of the approximating boundary B. Distance $d_2$ needs to be minimized as a function of $m_1$ and reaches its minimum when $m_1$ is midway between $n_1$ and $R^{-1}n_2$ and normalized to unit magnitude (this is well defined provided the distance $d_3$ is less than 1). The resulting unit normal vectors $m_1$ and $m_2$ for the approximating boundary B are such that the discrepancy between $n_i$ and $m_i$ (i.e. the distance $d_2$ in the inclination subspace) is minimized and the symmetry between grain 1 and grain 2 is preserved. This is equivalent to the geometrical construction shown in Fig. S2(b) where half of the residual rotation is applied to each grain to yield two new grain orientations, while orientation of the boundary plane in the lab-frame remains unchanged. The two correction angles approach $\pm d/2$ for small $d$.

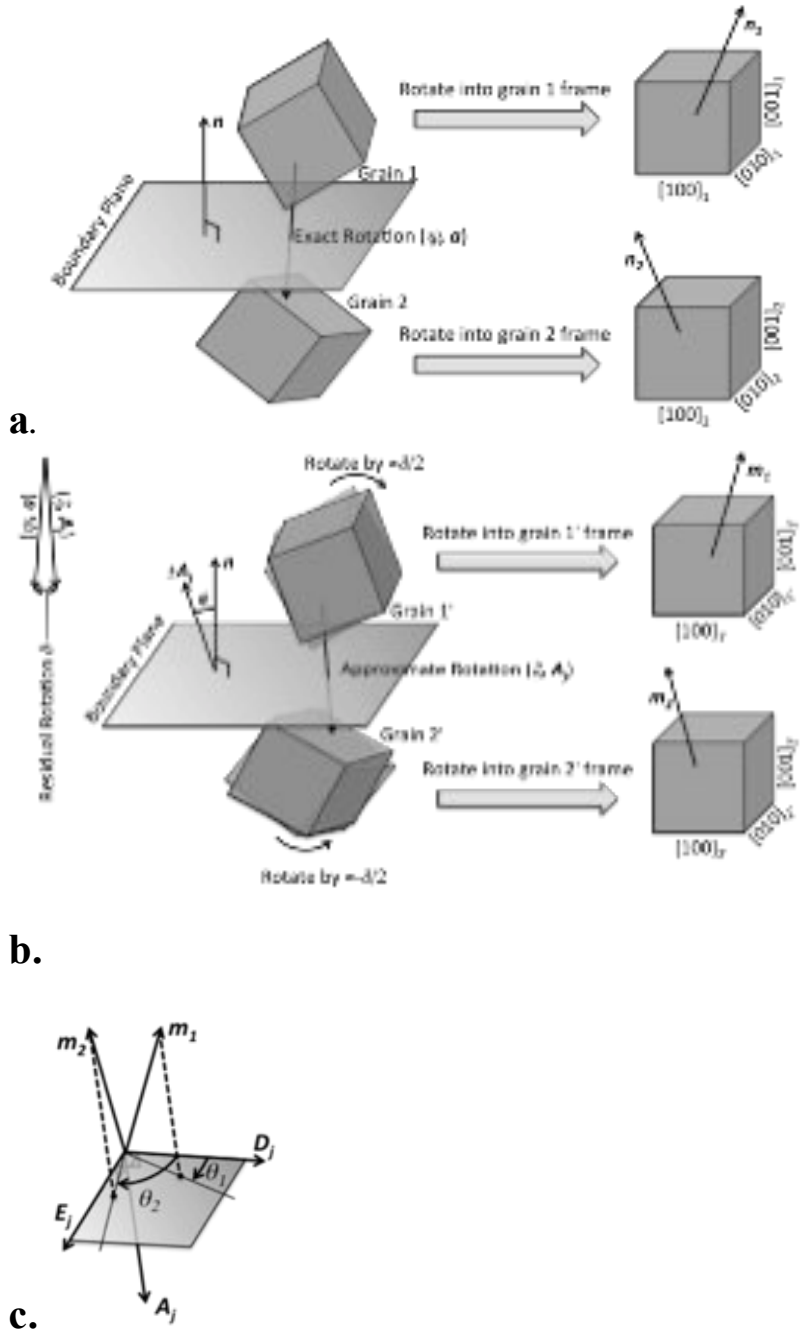

**Figure S2:** Definition of geometrical parameters. **a.** Parameters defining the grain boundary, including the axis-angle ($y$, $a$) of the rotation and the boundary plane unit normal in each grain's reference frame. **b.** Finding the best match for boundary A among all boundaries obtainable by rotation about some axis $A_j$. The closest match ($x$, $A_j$) to the exact misorientation ($y$, $a$) leaves a discrepancy that can be represented by a rotation by a small angle $d$. With the boundary plane fixed in the lab frame, each grain is rotated by approximately $\pm d/2$, defining new approximate grain orientations 1' and 2' defined by unit normals $m_1$ and $m_2$ expressed in the reference frames of these two grains. **c.** An orthonormal coordinate system ($A_j$, $D_j$, $E_j$) is defined in a crystal reference frame, e.g. ([100],[010],[001]), for each axis $A_j$. $D_j$ is always chosen to be a <100> or a <110> direction. The boundary plane normals in the approximated grain boundary are projected into the ($D_j$, $E_j$) plane, thus defining the angles $q_1$ and $q_2$.

For each of the 24 symmetry-related representations of boundary A and for each of 13 high symmetry rotation axes $A_j$ we find the best matching approximation $(x, A_j, m_1, m_2)$ and its misorientation distance $d_3$ to A. Resulting matches with $d_3 \geq d^{max}$ (with $d^{max}$ being different for <100>, <110>, and <111> rotations) are removed from the list. Also, for each special axis $A_j$ we remove redundant representations of A related to each other by the rotational symmetry of the axis. It is convenient to represent the information contained in $m_1$ and $m_2$ as angles rather than as unit vectors: we define the relevant angles as shown in in Fig. S2(b,c). Angle $f$ is nothing other than the angle, taken in the first quadrant, between the rotation axis $\pm A_j$ and the unit normals $m_1$ and $m_2$ of the approximating boundary, as shown in Fig. S2(b). Thus $f = 0$ for a twist boundary, $f = \pi/2$ for a tilt boundary, and $f$ takes on intermediate values for mixed boundaries. Angles $q_1$ and $q_2$ are defined as shown in Fig. S2(c), i.e. by projecting $m_1$ and $m_2$ onto a plane perpendicular to $A_j$. In order to define an origin and the sense of rotation for this plane, for each symmetry axis $A_j$ we define a crystallographic reference direction $D_j$ perpendicular to it; for example, for $A_j = [100]$ we set $D_j = [010]$. While in principle arbitrary, by convention $D_j$ is always taken to be an axis of at least two-fold rotational symmetry, hence $D_j$ is a <100> direction when $A_j$ is a <100> or <110> direction, and $D_j$ is a <110> direction when $A_j$ is a <111> direction. The third direction $E_j = A_j \times D_j$ completes an orthonormal coordinate system associated with each $A_j$ in which $q_1$ and $q_2$ are defined as shown. We recognize that the misorientation angle $x$ for the approximating boundary (Fig. S4(b)) is nothing other than $q_2 - q_1$ (except for the degenerate case of a pure twist boundary where $q_1$ and $q_2$ are undefined). We complete our specification of the boundary geometry by defining an asymmetry angle $h = q_2 + q_1$, which is zero for an STGB. By convention, $h$ is set to zero for a twist boundary.

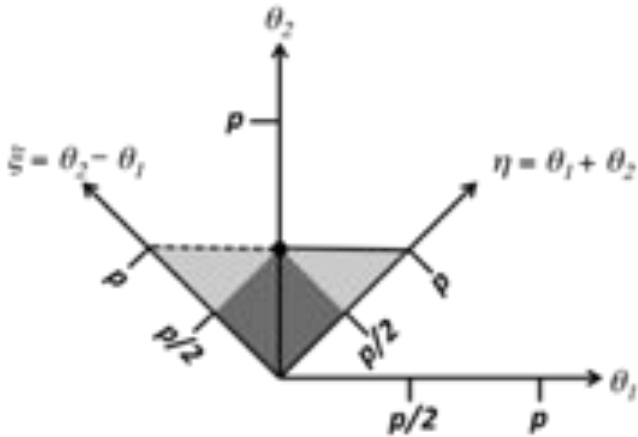

**Figure S3:** Transformation from $(q_1, q_2)$ space to $(x, h)$ space and reduction via symmetry to a semi-open right triangular region (union of light gray and dark gray regions). All functions in this space are periodic in $q_1$ and $q_2$ with period $p$ and possess mirror symmetry with respect to $x$ and $h$ axes. The point marked with a dot has 2-fold rotation symmetry, thus the boundary segment shown in a dashed line is redundant. All other boundary segments are part of the set. Twist boundaries are defined to have $h = 0$. While such symmetry reduction is applicable to all high-symmetry axes, the set of <111> tilt boundaries possesses an extra symmetry reducing the fundamental zone to the dark gray region.

To avoid redundancy, we map all points in the $(h, x)$ space to a minimal symmetry-equivalent zone as shown in Fig. S3. This remapping accounts for several symmetries. First, rotations about each axis $hkl$ are periodic with period $p_{hkl}$, being $\pi/2$, $2\pi/3$, and $\pi$ for <100>, <111>, and <110> rotations, respectively. Thus we can always map $q_1$ and $q_2$ into the semi-open interval $(-p_{hkl}/2, p_{hkl}/2]$. This compensates for the arbitrary choice we made in selecting a single $D_j$ for each $A_j$. Second, interchanging the labels on grains 1 and 2 (which, after accounting for the periodicity, would reverse the sign of $x$ but would not affect $h$) cannot change the physical character of the boundary. Thus we may redefine $x = |q_2 - q_1|$. Third, $D_j$ is always chosen to be a two-fold axis, and the FCC lattice itself has inversion symmetry, so that 180° rotations about the origin of this graph should have no physical effect. In combination with the other symmetries, this means that we can redefine $h = |q_2 + q_1|$. Furthermore, we can also insist that, whenever $q_2 = p_{hkl}/2$, $q_1$ can be replaced with $|q_1|$, thus creating the open edge segment at the upper left of the triangular region. <111> tilt boundaries have an additional symmetry because the 60° [111] rotation produces a physically equivalent result to a mirror reflection with respect to the (111) plane.

**2.3 Functional Forms**

With the geometrical definitions complete, we can now present the energy function in purely algebraic terms. The five-dimensional function is built hierarchically, starting from its one-dimensional subsets of pure twist and symmetric tilt boundaries, then expanding to the two-dimensional subsets of tilt boundaries, then to the three-dimensional scaffolding sets of <100>, <111> and <110> boundaries and finally to the entire 5-space.

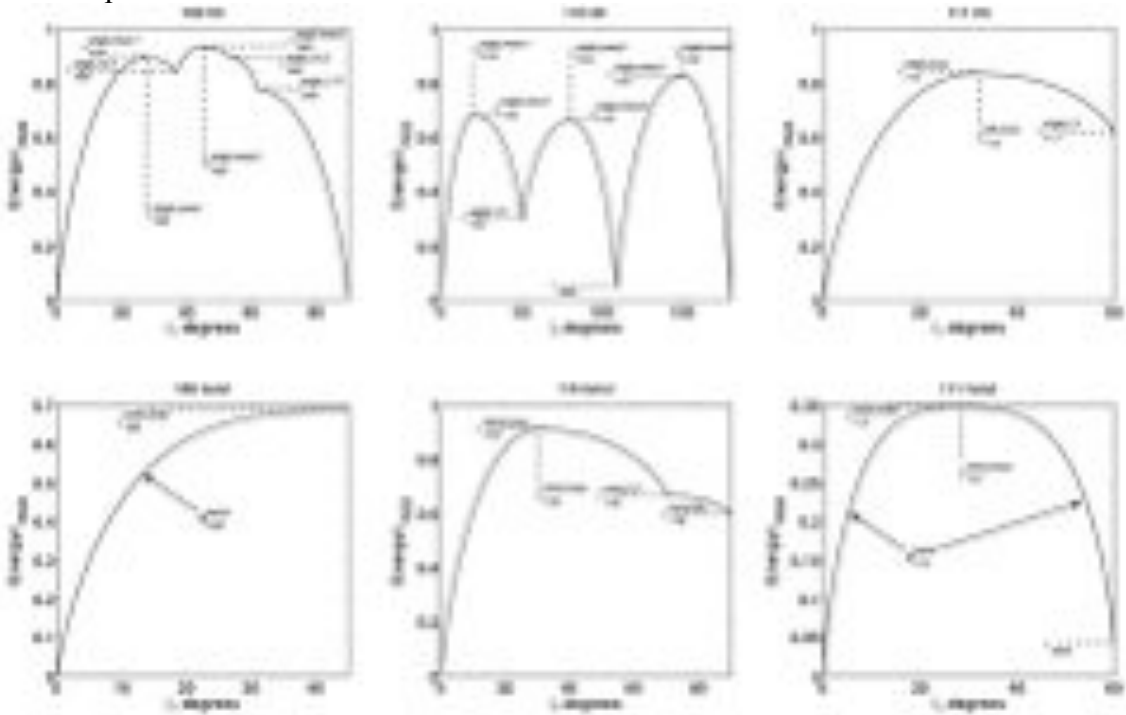

**Figure S4:** Functions describing GB energy variations within special one-dimensional subsets of twist and symmetric tilt boundaries in the <100>, <110>, and <111> scaffolding sets. Numerical parameters (angles and scaled energies) defining the shapes of the one-dimensional functions are listed on the plots. Unless shown on the plot, parameters *a* defining the shape of the RSW functions are set to *a* = 0.5.

To describe energy variations within the one-dimensional subsets we use the Read-Shockley-Wolf (RSW) function proposed by D. Wolf [s7] as an empirical extension of the classical Read-Shockley solution for the energy of the low-angle and vicinal boundaries. The RSW function is defined as

$$f_{RSW}(\theta; \theta_{min}, \theta_{max}, a) = \sin\left(\frac{\pi}{2}\frac{\theta-\theta_{min}}{\theta_{max}-\theta_{min}}\right)\left(1 - a \log \sin\left(\frac{\pi}{2}\frac{\theta-\theta_{min}}{\theta_{max}-\theta_{min}}\right)\right) \quad ,$$
(S1)

on segment [$q_{min}$, $q_{max}$] over which it ranges from 0 to 1, with a removable discontinuity at and an infinite slope at $q_{min}$ and a slope of zero at $q_{max}$. Within these limitations, the dimensionless shape parameter *a* can be used to modify the shape of the RSW function for various applications. For most of our curve fits we find *a* = 1/2 to be satisfactory. We use this latter value of *a* except where otherwise noted: the fitting results are rather insensitive to *a* in part because the set of 388 boundaries has very few vicinal boundaries and there is little information to more tightly constrain the values of *a*. The RSW function is convenient for approximating energy variations within one-dimensional subsets of pure twist and symmetric tilt boundaries within the <100>, <110>, and <111> scaffolding 3-sets. These six functions stitch together appropriate RSW functions as shown in Fig. S4. The figure also lists all the parameters defining the one-dimensional functions.

The two-dimensional subsets of all (symmetric + asymmetric) tilt boundaries are linearly interpolated between the one-dimensional symmetric-tilt subsets for <100> and <110>:

$$\epsilon_{100}^{tilt}(\xi,\eta) = \begin{cases} \epsilon_{100}^{stgb}\left(\frac{\pi}{2}-\xi\right) + \left(\epsilon_{100}^{stgb}(\xi) - \epsilon_{100}^{stgb}\left(\frac{\pi}{2}-\xi\right)\right)\left(1-\frac{2\eta}{\pi}\right)^p, \epsilon_{100}^{stgb}(\xi) > \epsilon_{100}^{stgb}\left(\frac{\pi}{2}-\xi\right) \\ \epsilon_{100}^{stgb}(\xi) + \left(\epsilon_{100}^{stgb}\left(\frac{\pi}{2}-\xi\right) - \epsilon_{100}^{stgb}(\xi)\right)\left(\frac{2\eta}{\pi}\right)^p, \text{otherwise} \end{cases}$$
(S2)

$$\epsilon_{110}^{tilt}(\xi,\eta) = \begin{cases} \epsilon_{110}^{stgb}\left(\frac{\pi}{2}-\xi\right) + \left(\epsilon_{110}^{stgb}(\xi) - \epsilon_{110}^{stgb}\left(\frac{\pi}{2}-\xi\right)\right) RSW\left(1-\frac{\eta}{\pi}, a_{110}^{atgb}\right), \epsilon_{110}^{stgb}(\xi) > \epsilon_{110}^{stgb}\left(\frac{\pi}{2}-\xi\right) \\ \epsilon_{110}^{stgb}(\xi) + \left(\epsilon_{110}^{stgb}\left(\frac{\pi}{2}-\xi\right) - \epsilon_{110}^{stgb}(\xi)\right) RSW\left(\frac{\eta}{\pi}, a_{110}^{atgb}\right), \text{otherwise} \end{cases}$$
. (S3)

This approach was found to be less effective for the two-dimensional set of <111> tilt boundaries that intersect the deep Σ3 grof. The latter grof appears to be well approximated by an RSW function with a cusp at *x* = π/3, however there is no evidence in

the available data for any *h* dependence of the <111> tilt energy for low values of *x*. Thus we define

$$\epsilon_{111}^{tilt}(\xi,\eta) = \begin{cases} \epsilon_{111}^{stgb,max} f_{R\ W}\left(\xi; 0, \xi_{111}^{tilt,max}, \frac{1}{2}\right), \xi \leq \xi_{111}^{tilt,max} \\ \epsilon_{111}^{stgb,\Sigma 3}(\eta) + (\epsilon_{111}^{stgb,max} - \epsilon_{111}^{stgb,\Sigma 3}(\eta))f_{RSW}\left(\xi; \frac{\pi}{3}, \xi_{111}^{tilt,max}, \frac{1}{2}\right), \text{otherwise} \end{cases}, \quad (S4)$$

where

$$\epsilon_{111}^{stgb,\Sigma 3}(\eta) = \epsilon_{111}^{stgb,\Sigma 3} + \left(\epsilon_{111}^{atgb,\Sigma 3,max} - \epsilon_{111}^{stgb,\Sigma 3}\right)f_{RSW}\left(\eta; 0, \frac{\pi}{2\lambda}, \frac{1}{2}\right) \quad (S5)$$

is the curve fit along the S3 grof. This introduces several additional fit parameters, including $\epsilon_{111}^{stgb,max}$, $\xi_{111}^{tilt,max}$, $\epsilon_{111}^{stgb,\Sigma 3}$, $\epsilon_{111}^{atgb,\Sigma 3,max}$, and a scaling parameter $\lambda$ that allows to account for the different shapes of the S3 grof in low-stacking-fault-energy and high-stacking-fault-energy materials. The value of $\lambda$ is unrestricted, leaving open the possibility of evaluating the RSW function outside of its natural domain.

The three-dimensional scaffolding sets of mixed - neither tilt nor twist - boundaries are built by interpolation between their twist and asymmetric tilt sub-sets as follows:

$$\epsilon_{hkl}(\xi,\eta,\phi) = \epsilon_{hkl}^{twist}(\xi)\left(1 - \frac{2\phi}{\pi}\right)^{p_{hkl}^1} + \epsilon_{hkl}^{tilt}(\xi,\eta)\left(\frac{2\phi}{pi}\right)^{p_{hkl}^2} \quad (S6)$$

for *hkl* = <100> and <110>, and as a parabola for *hkl* = <111>

$$\epsilon_{111}(\xi,\eta,\phi) = \epsilon_{111}^{twist}(\xi)\left(1 - \alpha\frac{2\phi}{\pi} + (\alpha - 1)\left(\frac{2\phi}{\pi}\right)^2\right) + \epsilon_{111}^{tilt}(\xi,\eta)\left(\alpha\frac{2\phi}{\pi} - (\alpha - 1)\left(\frac{2\phi}{\pi}\right)^2\right) \quad (S7)$$

The variables $p_{hkl}^1$, $p_{hkl}^2$, and *a* are fit parameters.

Finally, we approximate energy $\varepsilon$ of an arbitrary boundary in the 5-space as a weighted average over all approximating (best matching) boundaries found within the three scaffolding 3-sets:

$$\epsilon = \frac{1 + \sum w_{hkl}(d_3)\epsilon_{hkl}(\xi,\eta,\phi)}{1 + \sum w_{hkl}(d_3)} \epsilon_{RGB}, \quad (S8)$$

where $\epsilon_{RGB}$ is the energy of a hypothetical random boundary. To avoid multiple counting we remove symmetry-redundant representations of the same boundary. We also remove representations that are too distant to any of the three 3-sets ($d_3 > d^{max}$). The weighting function is

$$w_{hkl}(d_3) = \frac{w_{hkl}^0}{\sin\frac{\pi d_3}{2d_{hkl}^{max}}(1-\frac{1}{2}\log\sin\frac{\pi d_3}{2d_{hkl}^{max}})-1} \qquad (S9)$$

where *hkl* is one of <100>, <110>, or <111>. This weighting function is again of the RSW type and describes the energies of boundaries vicinal to one of the scaffolding sets, i.e. for which one of the distances $d_3$ is much smaller than the other distances. In order to keep the interpolation weights finite, we set $w_{hkl}(d_3)$ equal to some large but finite value when $d_3$ is numerically close to zero. We re-define all energy parameters in the $e_{hkl}(x, h, f)$ functions to be dimensionless expressing them as fractions of $\epsilon_{RGB}$ which defines the overall energy scale and is the only parameter with the dimension of energy.

## 2. 4. Recasting as a universal FCC function

In all, our function contains 43 numerical parameters - one dimensioned parameter ($\epsilon_{RGB}$) and 42 dimensionless parameters referred to as ***P*** (a 42-vector) – introduced to describe GB energy variation over five-dimensional space for a single element. These parameters are listed in Table S1 for all four elements, with parameters 2 through 43 making up vector ***P*** that encodes the *shape* of the function, as opposed to parameter 1 that encodes its *scale*. At first we used these 43 parameters to fit the sets of 388 boundary energies separately for each of four metals. This captured most of the reliable variance in the four data sets, reducing the residual RMS error to 5% (reduced $\chi^2$ ranging from 0.90 to 1.16 for the four elements with an assumed 5% error in all MD-calculated values). The resulting fits revealed two important relationships between the sets of best-fit parameters obtained for the four metals. First, 27 out of 42 components of the four best-fit vectors ***P*** obtained separately for the four metals were found to be very nearly the same to within a small error. Second, from the principal component analysis on the set of four ***P*** vectors we found that up to 97% of the inter-element variance in ***P*** could be captured with a single scalar parameter that describes the linearly coordinated motion of the remaining 15 elements of ***P***.

The above observations prompted us to redefine our interpolation function as universal. For that, 27 elements of ***P*** were deemed to be constant across the four elements. Furthermore, we elected to retain only one scalar parameter to describe the inter-element variations in ***P*** and to neglect all other principal components. Thus, our universal function contains only two material-dependent parameters: $\epsilon_{RGB}$ and one additional dimensionless parameter to account for the observed linearly coordinated variations in 15 out of 42 elements of ***P***. Given their linear dependence, any of the 15 parameters can be used as the master degree of freedom. To describe the inter-element variations in ***P*** we introduce a single coordinate $F$

$$\boldsymbol{P}(\Phi) = \boldsymbol{P}_{Al} + (\boldsymbol{P}_{Cu} - \boldsymbol{P}_{Al})\Phi . \qquad (S10)$$

As defined in (S10), parameter $F$ is zero for Al and 1 for Cu.

With the energy function re-defined as universal, a global search was performed to obtain the best-fitting set of numerical parameters using all four datasets containing 388x4 = 1552 boundary energies. In total, our universal energy function contains 63 parameters, including $\epsilon_{RGB}$ for all four elements, $\Phi_{Au}$, $\Phi_{Ni}$, $\boldsymbol{P}_{Al}$ (with 42 components), and the 15 elements of $\boldsymbol{P}_{Cu}$ that differed from the corresponding elements of $\boldsymbol{P}_{Al}$. The results of the global fit are shown in the main text and in figures S5-S7. Assuming again a 5% random error across all four atomistic datasets, the total value of $\chi^2$ increased by an insignificant amount compared to the values obtained in four independent fits (one for each element). At the same time, despite a significant reduction in the number of fitting parameters from 43x4 = 172 to 63, the reduced $\chi^2$ of 0.984 obtained in the global fit is even lower than the average reduced $\chi^2$ obtained in the four independent fits. This means that no statistically significant information is lost be describing all of the inter-element variation with just two parameters instead of 43.

Parameter $\Phi$ defines inter-element variations in the shape of the five-dimensional interpolation function and can be thought of as the element's position on a hypothetical aluminum-copper axis. This parameter correlates most strongly with the ratio of the coherent twin energy to $\epsilon_{RGB}$ (this ratio is shown as $\epsilon_{twin}$ in Table S1). Note that $\epsilon_{twin}$ appears twice in Fig. S4, since a coherent twin boundary is both a <110> tilt boundary and, simultaneously, a <111> twist boundary. As can be seen by comparing the curves for Al and Au, moving along the $F$ axis has also some more subtle effects, e.g. affecting the shape of the <111> twist, <111> tilt, and <100> twist sub-sets. The curve fit described below ultimately yielded values of 0.768 ± 0.022 and 0.784 ±0.022 for Ni and Au meaning that, apart from an overall scaling factor, the GB energy anisotropy for Ni and Au is nearly the same which could be expected, as an afterthought, given the narrow scatter of energy values for Au as seen in Fig. S1.

The set of 63 best-fit parameters obtained in the global fit is given in Table S1 along with their estimated errors derived using standard techniques [s8], namely from the square roots of the diagonal elements of the inverse of the $\chi^2$ curvature matrix. Such analyses can artificially inflate some of the parameter errors, especially in the (probably frequent) cases where the effects of changing one parameter can be partially compensated by changes in one or more other parameters. Thus, the errors shown in Table S1 should not be regarded as statistically independent.

compensated by changes in one or more other parameters. Thus, the errors shown in Table S1 should not be regarded as statistically independent.

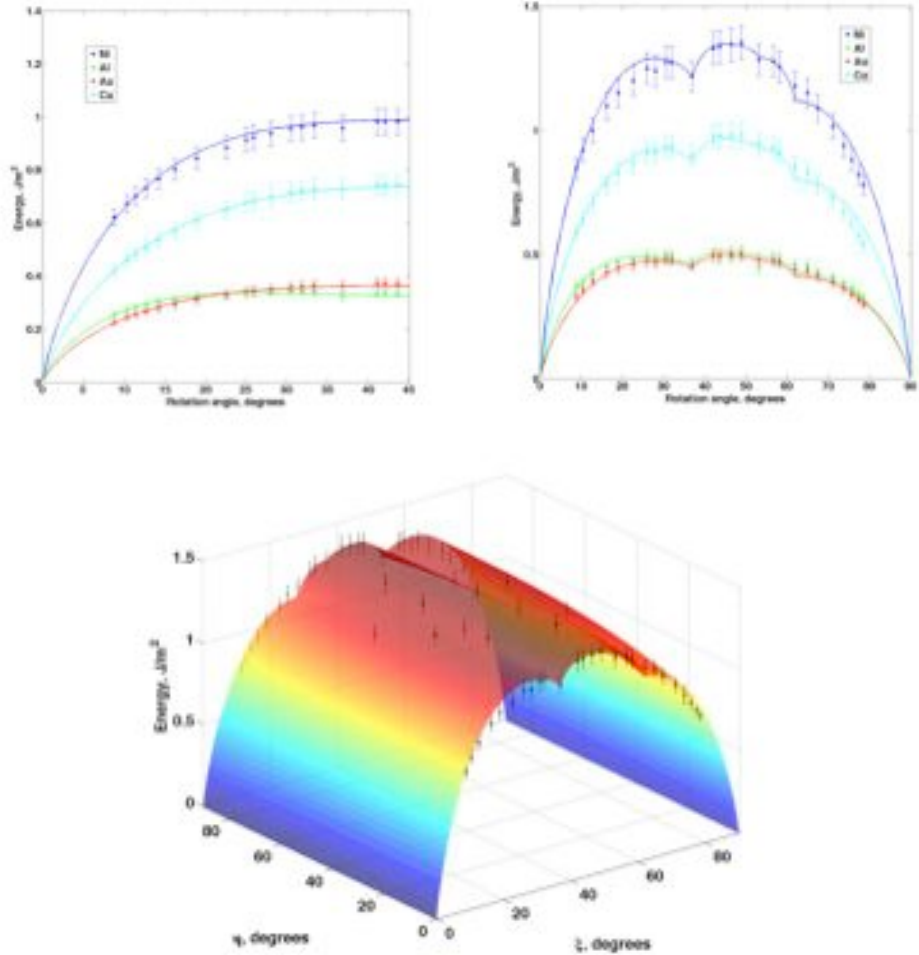

**Fig. S5:** GB energy variations in low-dimensional subsets within the 100 rotational set. The symbols are energies taken from the datasets of 388 boundaries with the assumed error bars of 5% while the lines and the surface show the fit function itself. (**A**) Energies of twist boundaries as a function of rotation angle. (**B**) Energies of symmetric tilt boundaries as a function of rotation angle. (**C**) Energies of all tilt (symmetric + asymmetric) boundaries as a function of rotation angle $\xi$ and asymmetry angle $\eta$ (for Ni only). Boundary character – twist, tilt or mixed - is defined by the orientation of the boundary plane relative to the rotation axis: twist/tilt boundaries have their planes perpendicular/parallel to the rotation axis and all other boundaries are considered mixed. Tilt boundaries are further classified as symmetric if the lattice planes parallel to the boundary are the same in both crystal grains and asymmetric otherwise. The lines at two borders of the surface at $\eta=0°$ and $\eta=90°$ in (C) are the same as the blue line in (B).

We make no claim of uniqueness of the function reported here. In its development we experimented with many minor variations of the functional form. What we report is the best result so far in terms of the total reduced $\chi^2$ across the set of four elements obtained with a function with just two element-specific parameters. However, several other functional forms (e.g. with RSW functions replaced with power laws or exponentials in some places, or using different weighted-average schemes (equation (S8))) produced results that were almost as good, with reduced $\chi^2$ values insignificantly larger than the value we report. In any case, the $\chi^2$ analysis shows that our functional form reduces the residual error to essentially random noise, assuming (as is consistent with the scatter plot shown in Fig. S1) that the calculated energies in the four datasets have a 5% RMS random error. Thus, while marginal improvements can still be made (for example, in the functional form for low-angle <110> STGBs, which were problematic for aluminum), significant further progress demands more data - either for additional FCC elements or for boundaries beyond the existing datasets of 388 boundaries. Of particular value would be additional data on the energies of vicinal boundaries and boundaries of very low symmetry that are poorly represented in the current ensemble.

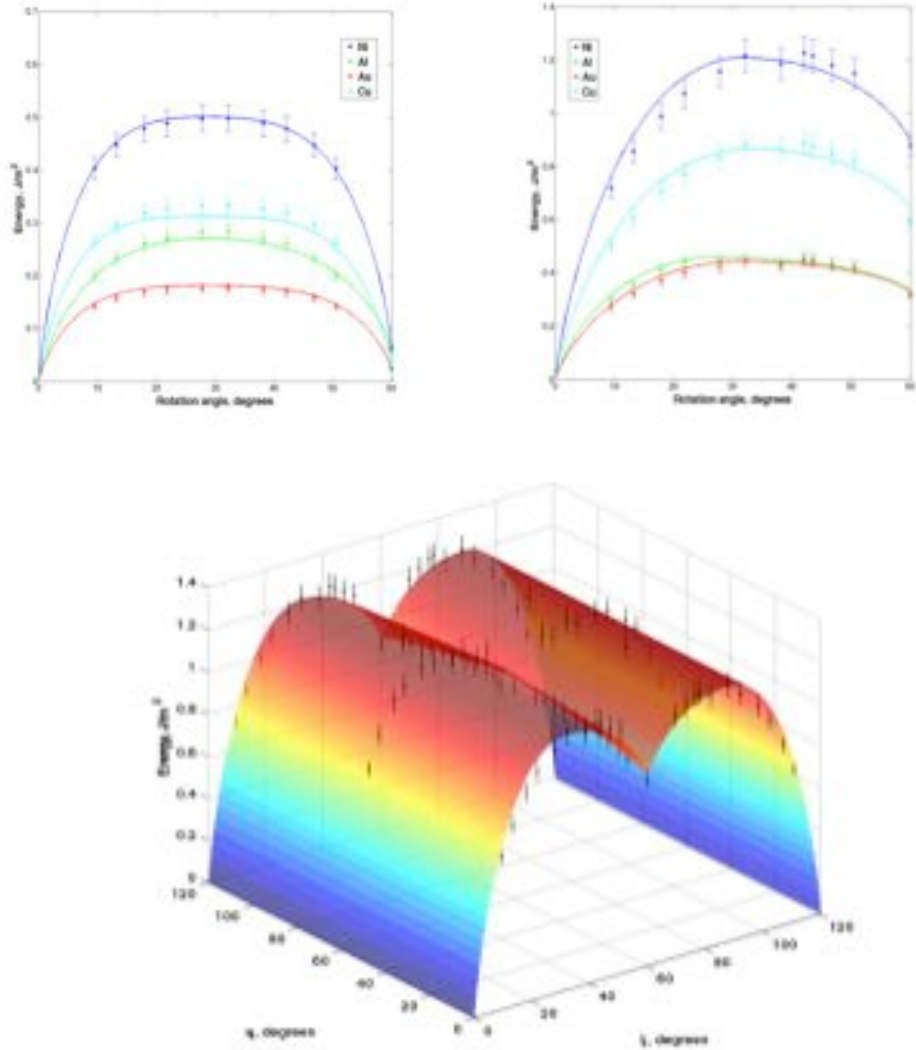

**Fig. S6:** GB energy variations in low-dimensional subsets within the 111 rotational set. The symbols are energies taken from the datasets of 388 boundaries with the assumed error bars of 5% while the lines and the surface show the fit function itself. (**A**) Energies of twist boundaries as a function of rotation angle. (**B**) Energies of symmetric tilt boundaries as a function of rotation angle. (**C**) Energies of all tilt (symmetric + asymmetric) boundaries as a function of rotation angle $\xi$ and asymmetry angle $\eta$ (for Ni only). Boundary character – twist, tilt or mixed - is defined by the orientation of the boundary plane relative to the rotation axis: twist/tilt boundaries have their planes perpendicular/parallel to the rotation axis and all other boundaries are considered mixed. Tilt boundaries are further classified as symmetric if the lattice planes parallel to the boundary are the same in both crystal grains and asymmetric otherwise. The lines at two borders of the surface at $\eta=0°$ and $\eta=120°$ in (C) are the same as the blue line in (B). This is the same as figure 2 in the text.

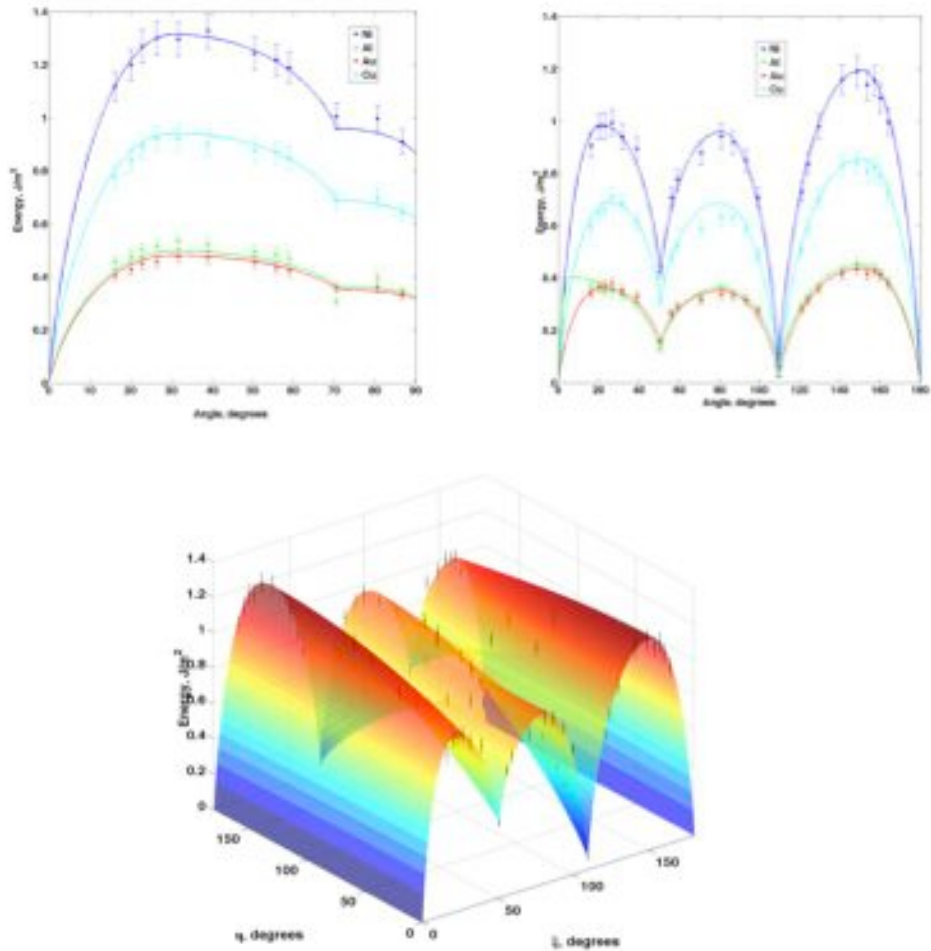

**Fig. S7:** GB energy variations in low-dimensional subsets within the 110 rotational set. The symbols are energies taken from the datasets of 388 boundaries with the assumed error bars of 5% while the lines and the surface show the fit function itself. (**A**) Energies of twist boundaries as a function of rotation angle. (**B**) Energies of symmetric tilt boundaries as a function of rotation angle. (**C**) Energies of all tilt (symmetric + asymmetric) boundaries as a function of rotation angle $\xi$ and asymmetry angle $\eta$ (for Ni only). Boundary character – twist, tilt or mixed - is defined by the orientation of the boundary plane relative to the rotation axis: twist/tilt boundaries have their planes perpendicular/parallel to the rotation axis and all other boundaries are considered mixed. Tilt boundaries are further classified as symmetric if the lattice planes parallel to the boundary are the same in both crystal grains and asymmetric otherwise. The lines at two borders of the surface at $\eta=0°$ and $\eta=180°$ in (C) are the same as the blue line in (B).

**Table S1:** All of the parameters for the global curve fit on all four elements. Parameter 1 is the over-all energy scale and is an independent parameter for each element. Parameters 2 through 43 constitute the dimensionless vector $P$. 27 of the elements of $P$ are held constant for all four elements, while the other 15 are taken to vary linearly with a single element-dependent parameter $\Phi$ (Equation S10). Uncertainties are included for the 63 independent parameters varied in the global fit.

| Number | Name | Description | Ni | Al | Au | Cu |
|---|---|---|---|---|---|---|
| 1 | $\varepsilon_{RGB}$ | Random boundary energy, J/m² | 1.445±0.032 | 0.547±0.012 | 0.530±0.012 | 1.037±0.023 |
| 2 | $d_{100}^{max}$ | Cutoff distance for 100 | colspan: 0.405±0.026 | | | |
| 3 | $d_{110}^{max}$ | Cutoff distance for 110 | colspan: 0.739±0.025 | | | |
| 4 | $d_{111}^{max}$ | Cutoff distance for 111 | colspan: 0.352±0.025 | | | |
| 5 | $w_{100}^{0}$ | Weight for 100 set | colspan: 2.40±0.60 | | | |
| 6 | $w_{110}^{0}$ | Weight for 110 set | colspan: 1.35±0.12 | | | |
| 7 | $w_{111}^{0}$ | Weight for 111 set | 2.676 | 0.352±0.250 | 2.726 | 3.38±0.85 |
| 8 | $p_{100}^{1}$ | 100 tilt/twist mix power law | colspan: 0.602±0.038 | | | |
| 9 | $p_{100}^{2}$ | 100 tilt/twist mix power law | colspan: 1.581±0.071 | | | |
| 10 | $\epsilon_{100}^{twist,max}$ | Maximum 100 twist energy | 0.684 | 0.596±0.017 | 0.686 | 0.710±0.018 |
| 11 | $a_{100}^{twist}$ | Shape factor for 100 twist | 0.871 | 1.310±0.064 | 0.861 | 0.738±0.037 |
| 12 | $p$ | 100 ATGB interpolation power law | colspan: 3.2±1.3 | | | |
| 13 | $\epsilon_{100}^{stgb,max1}$ | 100 STGB energy, first peak | colspan: 0.893±0.022 | | | |
| 14 | $\epsilon_{100}^{stgb,\Sigma5,1}$ | 100 STGB energy, first Σ5 | colspan: 0.835±0.024 | | | |
| 15 | $\epsilon_{100}^{stgb,max2}$ | 100 STGB energy, second peak | colspan: 0.933±0.026 | | | |
| 16 | $\epsilon_{100}^{stgb,\Sigma5,2}$ | 100 STGB energy, second Σ5 | colspan: 0.896±0.022 | | | |
| 17 | $\epsilon_{100}^{stgb,\Sigma17}$ | 100 STGB energy, Σ17 | colspan: 0.775±0.018 | | | |
| 18 | $\xi_{100}^{stgb,max1}$ | 100 STGB angle of first peak | 0.482 | 0.392±0.025 | 0.484 | 0.510±0.020 |
| 19 | $\xi_{100}^{stgb,max2}$ | 100 STGB angle of second peak | colspan: 0.783±0.114 | | | |
| 20 | $p_{110}^{1}$ | 110 tilt/twist mix power law | 0.743 | 0.679±0.025 | 0.744 | 0.762±0.024 |
| 21 | $p_{110}^{2}$ | 110 tilt/twist mix power law | 1.115 | 1.147±0.062 | 1.114 | 1.105±0.051 |
| 22 | $\xi_{110}^{twist,max}$ | 110 twist energy peak angle | colspan: 0.529±0.019 | | | |
| 23 | $\epsilon_{110}^{twist,max}$ | 110 twist energy peak value | colspan: 0.909±0.022 | | | |
| 24 | $\epsilon_{110}^{twist,\Sigma3}$ | 110 twist Σ3 | colspan: 0.664±0.020 | | | |
| 25 | $\epsilon_{110}^{twist,90}$ | 110 twist 90 degree (symmetry point) | colspan: 0.597±0.026 | | | |
| 26 | $a_{110}^{atgb}$ | 110 ATGB RSW shape factor | colspan: 0.200±0.055 | | | |
| 27 | $\epsilon_{110}^{stgb,max3}$ | 110 STGB energy of third peak | colspan: 0.826±0.018 | | | |
| 28 | $\epsilon_{twin}$ | 110 STGB energy Σ3 (coherent twin) | 0.043 | 0.111±0.005 | 0.042 | 0.0226±0.0012 |
| 29 | $\epsilon_{100}^{stgb,max2}$ | 110 STGB energy of second peak | colspan: 0.664±0.016 | | | |
| 30 | $\epsilon_{100}^{stgb,\Sigma11}$ | 110 STGB energy Σ11 | 0.285 | 0.242±0.012 | 0.286 | 0.298±0.011 |
| 31 | $\epsilon_{100}^{stgb,max1}$ | 110 STGB energy of first peak | 0.683 | 0.736±0.025 | 0.681 | 0.666±0.017 |
| 32 | $\pi - \xi_{110}^{stgb,max3}$ | 110 STGB position of third peak | colspan: 0.515±0.013 | | | |
| 33 | $\pi - \xi_{110}^{stgb,max2}$ | 110 STGB position of second peak | colspan: 1.738±0.020 | | | |
| 34 | $\pi - \xi_{110}^{stgb,max1}$ | 110 STGB position of first peak | 2.779 | 3.047±0.066 | 2.773 | 2.698±0.025 |
| 35 | $\alpha$ | 111 tilt-twist interpolation (linear if 1) | 1.851 | 1.490±0.124 | 1.858 | 1.960±0.059 |
| 36 | $a_{111}^{twist}$ | 111 twist RSW shape factor | 0.883 | 0.665±0.101 | 0.888 | 0.949±0.073 |

| 37 | $\xi_{111}^{twist,max}$ | 111 twist peak angle | \multicolumn{4}{c}{0.495±0.017} |
| --- | --- | --- | --- | --- | --- | --- |
| 38 | $\epsilon_{111}^{twist,max}$ | 111 twist energy at peak | 0.347 | 0.495±0.016 | 0.344 | 0.302±0.009 |
| 39 | $\xi_{111}^{tilt,max}$ | 111 ATGB peak angle | 0.550 | 0.469±0.021 | 0.551 | 0.574±0.015 |
| 40 | $\epsilon_{111}^{stgb,max}$ | 111 ATGB maximum energy | \multicolumn{4}{c}{0.837±0.019} |
| 41 | $\epsilon_{111}^{stgb,\Sigma 3}$ | 111 STGB Σ3 energy | \multicolumn{4}{c}{0.619±0.020} |
| 42 | $\epsilon_{111}^{atgb,\Sigma 3,max}$ | 111 ATGB Σ3 symmetry point energy | \multicolumn{4}{c}{0.845±0.143} |
| 43 | $\lambda$ | 111 ATGB mixing $\eta$ scale factor | 0.275 | 1.0±1.0 | 0.259 | 0.049±0.057 |
|  | $\Phi$ | Element-dependent shape factor | 0.768±0.022 | 0 | 0.784±0.022 | 1 |

## 3. Description of the MATLAB code

On xxx.llnl.gov we post a set of MATLAB functions, in the form of '.m' files, that calculate the energy of any given grain boundary in one of the four FCC metals Ni, Al, Au, Cu or in any other user-defined FCC material. The boundary should be represented as a pair of orthonormal rotation matrices *P* and *Q* whose rows specify orientation of the laboratory (sample) frame in the cube coordinates of crystal grains 1 and 2, respectively. The top row in each matrix defines the orientation of the boundary plane normal written in the two cube frames. The calling sequence for computing the interpolated GB energy should be, for example:

```
[geom100,geom110,geom111] = distances_to_all_sets(P,Q);   % Generate geometry parameters
par43 = makeparvec('Ni');                                  % Generate 43-element parameter vector for nickel
en = weightedmeanenergy(geom100,geom110,geom111,par43);    % Calculate the energy
```

The function 'makeparvec' contains four parameters that by default will revert to the values obtained by curve fitting to the full set of 4 elements (i.e. the values in Table S1). The first parameter, 'AlCuparameter', is either a numeric scalar (in which case it is interpreted as the parameter $\Phi$ in equation (S10)) or a character string (in which case it is interpreted as the symbol for one of the four elements Ni, Al, Au, or Cu). The second parameter is the energy of random boundary $e_{RGB}$ in J/m$^2$ and defaults to the value for copper. The third and fourth parameters are the 42-element vectors $\boldsymbol{P_{Al}}$ and $\boldsymbol{P_{Cu}}$, also from equation (S10). Calling 'makeparvec' with no parameters returns the parameter set for copper.

The above functions call several other functions also supplied on the website: 'distances_to_all_sets' calls 'distances_to_set' that calculates the geometrical parameters for one of the sets <100>, <110>, and <111>. 'weightedmeanenergy' calls functions with specific curve fits to each set, called 'set100', 'set110', and 'set111', and these in turn call the general-use function 'rsw'.